\documentclass[twocolumn,superscriptaddress,showpacs,prd,aps,amsmath,amssymb,nofootinbib]{revtex4-1}
\usepackage{graphicx}
\usepackage{bm} % bold math
\usepackage{mathrsfs} % -> \mathscr
\usepackage{color}

\newcommand{\Madm}{M_{\rm ADM}}
\newcommand{\MK}{M_{\rm K}}

\newcommand{\beq}{\begin{equation}} 
\newcommand{\eeq}{\end{equation}} 
\newcommand{\beqn}{\begin{eqnarray}} 
\newcommand{\eeqn}{\end{eqnarray}} 
\newcommand{\pa}{\partial}
\newcommand{\na}{\nabla}

\newcommand{\gabd}{g_{\alpha\beta}}

\newcommand{\gmabd}{\gamma_{ab}}
\newcommand{\tgmabu}{\tilde\gamma^{ab}}
\newcommand{\tgmabd}{\tilde\gamma_{ab}}
\newcommand{\tgamma}{\tilde\gamma}

\newcommand{\fabd}{f_{ab}}

\newcommand{\Aabu}{A^{ab}}
\newcommand{\Aabd}{A_{ab}}

\newcommand{\albe}{{\alpha\beta}}

\newcommand{\Fabd}{F_{\alpha\beta}}

\newcommand{\ttR}{{}^{3}\!\tilde R}

\newcommand{\tD}{\tilde D}
\newcommand{\zD}{{\raise1.0ex\hbox{${}^{\ \circ}$}}\!\!\!\!\!D}
\newcommand{\alone}{{\raise0.5ex\hbox{${}^{\ 1}$}}\!\!\!\!\alpha}
\newcommand{\Od}{{O}}

\newcommand{\Lie}{\mbox{\pounds}}

\newcommand{\nalam}{\mathrel{\raise0.9ex\hbox{$^\lambda$}\mkern-14mu
\lower0.0ex\hbox{$\nabla$}}}

\newcommand{\dis}{\displaystyle}
\newcommand{\gmaa}{\gamma_a\!{}^\alpha}
\newcommand{\gmbb}{\gamma_b{}^\beta}

\newcommand{\zeroD}{{\raise1.0ex\hbox{${}^{\ \circ}$}}\!\!\!\!\!D}

\newcommand{\zLap}{{\raise1.0ex\hbox{${}^{\ \circ}$}}\!\!\!\!\Delta}
\newcommand{\zna}{{\raise1.0ex\hbox{${}^{\ \circ}$}}\!\!\!\!\!\nabla}
\newcommand{\zS}{{\raise1.0ex\hbox{${}^{\ \circ}$}}\!\!\!\!\!S}

\newcommand{\hu}{h{\underbar u}}

\newcommand{\cocal}{{\sc cocal }}

\begin{document}

\title{
Equilibrium solutions of relativistic rotating stars 
with mixed poloidal and toroidal magnetic fields}

\author{K\=oji Ury\=u}
\email{uryu@sci.u-ryukyu.ac.jp}
\affiliation{
Department of Physics, University of the Ryukyus, Senbaru 1, 
Nishihara, Okinawa 903-0213, Japan}
\author{Eric Gourgoulhon}
\email{eric.gourgoulhon@obspm.fr}
\affiliation{
Laboratoire Univers et Th\'eories, UMR 8102 du CNRS,
Observatoire de Paris, Universit\'e Paris Diderot, F-92190 Meudon, France}
\author{Charalampos M.~Markakis} 
\email{charalampos.markakis@uni-jena.de}
\affiliation{
Theoretical Physics Institute,  University of Jena, Max-Wien-Platz 1, 07743 Jena, Germany}
\affiliation{
 Mathematical Sciences, University of Southampton, Southampton, SO17 1BJ, United Kingdom}
\author{Kotaro Fujisawa}
%\email{fujisawa@ea.c.u-tokyo.ac.jp}
\email{fujisawa@heap.phys.waseda.ac.jp}
\affiliation{
Advanced Research Institute for Science and Engineering, Waseda University, 3-4-1 Okubo, 
Shinjuku-ku, Tokyo 169-8555, Japan}
\author{Antonios Tsokaros}
\email{tsokaros@th.physik.uni-frankfurt.de}
\affiliation{
Institut f\"ur Theoretische Physik, 
Johann Wolfgang Goethe-Universit\"at, 
Max-von-Laue-Str. 1, 
60438 Frankfurt am Main, Germany 
}
\author{Yoshiharu Eriguchi}
\email{eriguchi@ea.c.u-tokyo.ac.jp}
\affiliation{
Department of Earth Science and Astronomy, Graduate School of Arts and Sciences, University of Tokyo, 
Komaba 3-8-1, Meguro, 153-8902 Tokyo, Japan}

\date{\today}  

\begin{abstract} 
%Fully general relativistic solutions of strongly magnetized rapidly rotating 
%stars are obtained.  
Stationary and axisymmetric solutions of relativistic rotating stars 
with strong mixed poloidal and toroidal magnetic fields 
are obtained numerically.  
Because of the mixed components of the magnetic field, the underlying stationary 
and axisymmetric spacetimes are no longer circular.  
These configurations are computed from the full set of the Einstein-Maxwell 
equations, Maxwell's equations and from first integrals and integrability 
conditions of the magnetohydrodynamic equilibrium equations. After a brief 
introduction of the formulation of the problem, 
we present the first results for highly deformed magnetized rotating
compact stars. 
%
% We solve such non-circular spacetimes using waveless formulation.  
%Computations for such solutions is a first step to develop models 
%for new born magnetars.  
\end{abstract} 

%\pacs{04.20.-q, 04.40.Dg, 04.40.Nr, 52.30.Cv, 95.30.Qd}

\maketitle
 
%\section{Introduction}
\paragraph*{\underline{Introduction}:}
Neutron stars (NS) with strong 
surface magnetic fields around $10^{14}-10^{15}$G 
are considered as the source of soft gamma 
repeaters and anomalous x-ray pulsars 
\cite{Duncan:1992hi}.  The widely accepted 
magnetar model stimulated an interest in 
constructing solutions of strongly magnetized 
relativistic rotating stars.  Numerical computations 
of such magnetized stars are useful tools for 
investigating allowed configurations of the interior 
magnetic fields, or a limit of the strength of the fields.  
In particular, compact stars with stronger magnetic fields 
could model newly born magnetars (see, e.g.\cite{Metzger:2010pp}).

An approach for studying magnetized rotating neutron stars is 
to obtain their equilibrium configurations.  
Stationary and axisymmetric solutions of rotating relativistic stars 
with strong poloidal fields have been successfully 
calculated in \cite{Bocquet:1995je}, and with toroidal fields 
in \cite{Kiuchi:2008ch,Frieben:2012dz}.  In these computations, 
the spacetime is assumed to be orthogonally transitive (circular -- 
invariant under a simultaneous inversion of $t\rightarrow -t$, 
and $\phi\rightarrow -\phi$), 
because the configuration of magnetic fields 
is restricted to either purely poloidal or toroidal fields 
and the velocity field is to circular 
\cite{Carter73,Oron:2002gs,2013rrs..book.....F,2013gere.book.....S} 
%(see, Appendix \ref{app:spacetime}).  %\cite{circular}.  
A relativistic formulation for more general magnetized relativistic 
stars with mixed poloidal and toroidal fields has been derived 
by Bekenstein and Oron \cite{1978PhRvD..18.1809B}.  
The formulation has been used to obtain slowly rotating and 
weakly magnetized solutions \cite{IokaSasaki,Ciolfi}, and has been improved 
for obtaining more stable solutions \cite{Yoshida:2012mu}.  
More recently, numerical solutions for mixed poloidal and toroidal fields 
have been obtained under a simplified relativistic gravity\cite{Pili:2014npa}.\footnote{
They have assumed the form of spacetime metric to be 
$ds^2 = -\alpha^2 dt^2 + \psi^4 f_{ij} dx^i dx^j$, where 
$f_{ij}$ is a flat spatial metric.
}

In this paper, we present a formulation and numerical solutions 
for rapidly rotating relativistic stars with 
strong mixed poloidal and toroidal magnetic fields.  Our solutions extend 
the above works \cite{IokaSasaki,Ciolfi,Yoshida:2012mu,Pili:2014npa} to highly deformed 
configurations as a result of rapid rotation and stronger magnetic fields, 
which can not be calculated from perturbative methods.  
We assume stationarity and axisymmetry of the gravitational, electromagnetic and matter fields, 
but we do not assume the spacetime to be circular, or spatially conformally flat.  
Moreover our 
new method of solving Einstein's and Maxwell's equations are not restricted 
to axial symmetry.  The first integrals of 
a system of relativistic magnetohydrodynamic (MHD) equations are derived 
assuming the above symmetries as well as perfect conductor conditions 
\cite{1978PhRvD..18.1809B,IokaSasaki,Gourgoulhon:2011gz}, 
and solved to obtain self-consistent, non-perturbative, equilibrium configurations for the first time.  

The existence and stability of mixed poloidal and toroidal magnetic fields in  
compact stars is supported by a number of recent simulations 
of magnetized rotating stars \cite{Braithwaite}.  
It is demonstrated that such stable configurations of mixed fields are 
reached from arbitrary initial data.  Several groups have been 
developing relativistic MHD simulation codes, 
and have performed some, rather preliminary, simulations of magnetized 
compact stars \cite{GRsimulations}.  Our equilibrium solutions 
provide initial data for such simulations aiming to obtain stable 
configurations of strongly magnetized compact stars.  
The stability of  equilibrium solutions may be determined 
from approaches complementary to the simulations, such as an 
application of a linear perturbation method described in 
\cite{Yoshida:2012rk}.  

In this paper, we use 
$G=c=4\pi\epsilon_0=1$ units unless otherwise specified.  Greek 
indices, $\alpha, \beta, ...$ are used for spacetime 
tensors, lower case latin indices, $a, b,... $ for 
3-dimensional spatial tensors, and upper case Latin indices, 
$A, B,... $ for 2-dimensional spatial tensors in a meridional plane.  
The exterior derivative of a 1-form $w_\alpha$ 
is denoted by $(dw)_\albe = \na_\alpha w_\beta - \na_\beta w_\alpha$.

%%% \section{Formulation}
\paragraph*{\underline{Formulation}:}
%
%%% \subsection{Field equations}
%
A relativistic rotating star with both poloidal and toroidal 
magnetic fields (and possibly with electric fields and/or meridional flows) 
can be modeled by Einstein-Maxwell charged  magnetized perfect-fluid spacetimes.  
Although stationarity and axisymmetry is assumed for spacetime, 
electromagnetic fields and matter, such spacetimes cannot be circular.\footnote{
%(see Appendix \ref{app:spacetime}).  
Stationary and axisymmetric spacetime is called circular if the 
2-planes orthogonal to two killing vectors $t^\alpha$ and $\phi^\alpha$ 
is integrable.  The derivation of the metric for circular spacetime is found 
in \cite{Carter73,2013rrs..book.....F,2013gere.book.....S,
Papapetrou,Kundt,1992mtbh.book.....C,1984ucp..book.....W}.  
It relies on the vanishing twists of 
two killing vectors $t^\alpha$ and $\phi^\alpha$, 
$ t_{[\alpha} \phi_\beta \na_\gamma t_{\delta]}\,=\,0$ and 
$t_{[\alpha} \phi_\beta \na_\gamma \phi_{\delta]}\,=\,0$ (Frobenius conditions), 
which are no longer satisfied for the spacetime associated 
with mixed poloidal and toroidal fields \cite{Oron:2002gs,supp1}.  
}
Such non-circular spacetimes have been calculated in the slow rotation and weak magnetic field limit 
as in \cite{IokaSasaki,Ciolfi,Yoshida:2012mu}, and in the fully nonlinear regime 
for non-magnetized compact star with large meridional flow in \cite{Birkl:2010hc}.  
A proposed formulation for computing non-circular spacetimes invokes a 2+1+1 spacetime decomposition 
 \cite{1993PhRvD..48.2635G,Birkl:2010hc}. Here, instead of this formulation, we 
use the one developed in \cite{Shibata:2004qz}, 
which is based on a 3+1 decomposition and has been successfully applied for computing waveless 
initial data for binary neutron stars (BNS)  \cite{WLBNS}.

The spacetime $({\cal M}, \gabd )$, ${\cal M}=\mathbb{R}\times\Sigma$, 
is foliated by a family of spacelike hypersurfaces $\Sigma_t = \chi_t(\Sigma_0)$ 
where the hypersurface $\Sigma=\Sigma_0$ is an initial slice.  
The diffeomorphism $\chi_t$ is generated by the time symmetry vector $t^\alpha$, 
so $\Sigma_t$ are identical for any $t$. 
The timelike vector $t^\alpha$ is tangent to the curves 
$t \rightarrow (t,x)\in \mathbb{R}\times\Sigma$, and is 
related to the future-pointing normal $n^\alpha$ of $\Sigma_t$ by 
$t^\alpha = \alpha n^\alpha + \beta^\alpha$, where $\alpha$ is the lapse 
and $\beta^\alpha$ the shift satisfying $\beta^\alpha n_\alpha=0$.  
The projection tensor $\gamma_\albe = \gabd+n_\alpha n_\beta$
restricted to a slice $\Sigma_t$ is equal to a spatial metric $\gmabd(t)$ 
on $\Sigma_t$.  We introduce a conformally rescaled spatial metric 
$\tgmabd = \psi^{-4}\gmabd$ whose decomposition is specified by 
the condition $\det \tgmabd = \det \fabd$, where $\fabd$ is a flat metric
on $\Sigma_t$.  
Then, the metric $\gabd$ in 3+1 form becomes 
$ds^2 = -\alpha^2 dt^2 + \psi^{4}\tgmabd(dx^a+\beta^a dt)(dx^b+\beta^b dt)$.  

The metric potentials $\{\psi, \beta_a, \alpha, \tgmabd \}$ are solved from 
the respective  components of the 3+1 Einstein-Maxwell equations, 
which consist of the Hamiltonian and momentum 
constraints, the spatial trace  part of Einstein-Maxwell equations, and those of the 
spatial tracefree parts, respectively (see e.g. \cite{Gourg2012,Bonazzola:2003dm,Shibata:2004qz}).  
As for coordinate conditions, maximal slicing $K=0$ and the spatially transverse condition 
(Dirac gauge) $\zD_b \tgmabu =0$ are imposed, where 
$K$ is the trace of the extrinsic curvature 
%% $K_{ab} := -\frac1{2\alpha} \Lie_n \gmabd$, and 
$K_{ab} := -\frac1{2\alpha} (\pa_t - \Lie_\beta)\gmabd$ of a slice $\Sigma_t$.  
$\zD_a$ is the covariant derivative with respect to the flat metric $\fabd$, and 
$\Lie_\beta$ is the Lie derivative along the shift $\beta^a$ defined on $\Sigma_t$.  
As discussed in \cite{Gourg2012,Bonazzola:2003dm,Shibata:2004qz}, 
all metric components have Coulomb type
fall off if the condition $\pa_t\tgmabu=\Od(r^{-3})$ is imposed.  
For the present case, the stationarity condition $\pa_t\tgmabu=0$ is 
imposed which is the same condition used in our previous computations 
for BNS data \cite{WLBNS}.  Finally, we solve for the Cartesian components of the spatial 
vector and tensor equations for 
$\beta_a$ and $\tgmabd$, which  are written as 3-dimensional Poisson equations with 
nonlinear sources. 

An analogous formulation is applied for solving Maxwell's equations.  
The electromagnetic 1-form $A_\alpha$ is decomposed in 3+1 variables 
$\Phi_\Sigma = -A_\alpha n^\alpha$, and $A_a = \gmaa A_\alpha$, 
and the projections of Maxwell's equations onto $\Sigma_t$ as well as 
along  the normal $n^\alpha$ to $\Sigma_t$ are written in terms of 
3+1 variables $\{\Phi_\Sigma, A_a\}$.  Imposing the stationarity condition 
$\Lie_t A_\alpha =0$,\footnote
{The Lie derivative $\Lie_t$ along the vector $t^\alpha$ 
is defined on $\cal M$.}
and the Coulomb gauge condition $\zD^a A_a=0$, 
Maxwell's equations reduce to Poisson equations for
$\{\Phi_\Sigma, A_a\}$ with nonlinear sources.

As mentioned above, gauge conditions for the metric and the electromagnetic 
fields consist of the maximal slicing $K=0$, the generalized Dirac gauge condition 
$\zD_b \tgmabu =0$, and the Coulomb gauge condition $\zD^a A_a=0$.  The condition 
$K=0$ can be imposed explicitly by replacing $K_{ab}$ with its tracefree part 
$A_{ab} = K_{ab} - \frac13\gmabd K$ in the above equations.  
To impose the other two gauge conditions, we introduce 
the gauge potentials $\xi^a$ for the Dirac gauge and $\xi$ for the Coulomb gauge, 
with which the field variables are transformed as 
$\tilde{\gamma}^{ab}{}' = \tgmabu - \zD^a \xi^b - \zD^b\xi^a+\frac23 f^{ab}\zD_c \xi^c$ and 
$A_a' = A_a - \zD_a \xi$, and let $\tilde{\gamma}^{ab}{}'$ and $A_a'$ satisfy 
the gauge conditions \cite{WLBNS}.  For the case of Coulomb gauge, a substitution of 
$A_a'$ to the gauge condition yields an elliptic equation for the gauge potential 
$\xi$, $\zD_a \zD^a \xi = \zD^a A_a$.  This equation and an analogous one for $\xi^a$ 
are solved simultaneously with the field equations at each cycle of iteration, and then 
the fields $\tgmabu$ and $A_a$ are modified according to the above gauge transformations.\footnote
{In the formulation \cite{Gourg2012,Bonazzola:2003dm}, 
the Dirac gauge condition is imposed on Einstein's equation explicitly 
to isolate two independent variables to be solved from dynamic equations.}
%All components of the metric $\gabd$ and the electromagnetic fields 
%$A_\alpha$ and gauge potentials 
%Because of the gauge freedom for the metric and the electroamgnetic field, 

%%% \subsection{Ideal MHD equations and electric currents}

The ideal MHD system of equations  includes the MHD-Euler equation, 
the rest mass conservation equation, and the perfect conductivity 
condition.  The numerical integration of these partial differential 
equations, from which time derivative terms are eliminated
in order to obtain equilibrium solutions, is technically challenging.  
Under the assumptions of stationarity and axisymmetry, however, 
one can find a set of integrability conditions and algebraic relations 
that the fluid variables must satisfy.  
Such a formulation for the system of relativistic ideal MHD equations 
is found in \cite{1978PhRvD..18.1809B,IokaSasaki,Yoshida:2012mu}, and 
a fully covariant geometric formulation is derived in our previous work 
\cite{Gourgoulhon:2011gz}.  
We rewrite the formulation in 
\cite{1978PhRvD..18.1809B,IokaSasaki,Yoshida:2012mu} as
suited to our numerical method.  
We introduce the basis, 
$\{t^\alpha,\phi^\alpha, e_A^\alpha\}$, associated with 
the coordinates $t$, $\phi$ and two other spatial coordinates
$x^A$, where $t$ and $\phi$ are chosen along the symmetry 
vectors $t^\alpha$ and $\phi^\alpha$ and normalized as 
$t^\alpha \na_\alpha t = 1$, $\phi^\alpha \na_\alpha \phi = 1$, and 
$e_A^\alpha \na_\alpha x^B = \delta_A{}^B$, where $\delta_A^B$ 
is the Kronecker delta.  
For example, the electric current density $j^\alpha$ expanded on
this basis is denoted by 
$j^\alpha = j^t t^\alpha +  j^\phi \phi^\alpha +  j^A e_A^\alpha$, and 
the 4-velocity by 
$u^\alpha = u^t t^\alpha +  u^\phi \phi^\alpha +  u^A e_A^\alpha$.  
We also introduce the projection tensor
$\sigma_a{}^b=\gamma_a{}^b-\phi^b\na_a\phi$: any
spatial tensor is projected onto a $\phi=\rm{constant}$ surface via $\sigma_A{}^{a}$.

For a relativistic magnetized (and charged) perfect 
fluid with infinite conductivity, the flow 
always becomes isentropic.  
From the isentropic flow condition, the rest mass conservation equation, and 
the projection of the perfect conductivity conditions along the symmetry vectors 
$t^\alpha$ and $\phi^\alpha$, we derive 
conditions for the entropy $S$, the weighted stream function of  meridional flow 
$\sqrt{-g}\Psi$, and the $t$ and $\phi$ components of the electromagnetic potential
1-form $A_\alpha$, to be functions of a master potential 
$\Upsilon$, $S = S(\Upsilon)$,
$\sqrt{-g}\Psi = [\sqrt{-g}\Psi](\Upsilon)$,\footnote
{See Eq.~(\ref{eq:stream_fn}) for a definition of the stream function for the meridional 
velocity fields $u^A$.}
$A_t = A_t(\Upsilon)$ and 
$A_\phi = A_\phi(\Upsilon)$, which will be referred to as integrability conditions.\footnote
{The integrability itself does not constrain the relation between each quantity of 
$S, \sqrt{-g}\Psi, A_t, A_\phi$, and the master potential $\Upsilon$, although 
a solution may not always exist for a certain relations that chosen arbitrarily, 
or a certain range of parameters involved in the relations.} 
The remaining meridional components of the perfect conductivity condition and 
MHD-Euler equations (together with the normalization of the 4-velocity) yield a set of 
algebraic relations that we call 
first integrals.  As a consequence of the two symmetries, any exact 2 form does not 
have $t\phi$-component.

A common strategy for computing stationary  axisymmetric equilibria 
of the matter and electromagnetic fields under the above integrability conditions 
is to derive the master equation, called transfield equation, for the master 
potential $\Upsilon$ by eliminating redundant variables from the first integrals 
and the Maxwell equations.  The transfield equation may be further 
simplified to a well-known Grad-Shafranov equation by imposing 
$\Upsilon = A_\phi$ and $\sqrt{-g}\Psi ={\rm constant}$.  
Either relation is an elliptic equation, defined on a $\phi ={\rm constant}$ plane, to be 
solved for the potential $\Upsilon$ or $A_\phi$ (see e.g. 
\cite{1978PhRvD..18.1809B,IokaSasaki,Yoshida:2012mu,Gourgoulhon:2011gz}).  

In our formulation, we do not solve such a master equation, 
but instead solve the first integrals and Maxwell's equations simultaneously.  
For the sources of Maxwell's equations, we rewrite the current density 
$j^\alpha$ using the first integrals derived from the MHD-Euler equation, 
hereafter assuming $\Upsilon=A_\phi$ for simplicity, 
\beqn
&&
j^A \sqrt{-g} 
\,=\,\left([\sqrt{-g}\Psi]''hu_\phi
+[\sqrt{-g}\Lambda_\phi]'
\right)\delta^{AB}B_B
\nonumber\\
&&\qquad\qquad
\,-\,[\sqrt{-g}\Psi]'
\delta^{AB}\omega_B,  
\label{eq:current_jA}
\\
&&
j^\phi\sqrt{-g}
\,+\,A'_t\, j^t\sqrt{-g}
\,=\,
\left([\sqrt{-g}\Psi]''hu_\phi 
+ [\sqrt{-g}\Lambda_\phi]'\right)B_\phi
\nonumber\\
&&\qquad\qquad
\,-\,[\sqrt{-g}\Psi]'\omega_\phi
\,-\,\left(A''_t \,hu_\phi+\Lambda'\right)\rho u^t \sqrt{-g}, 
\label{eq:current_jphijt_Aphi}
\eeqn
where 
$B_B$ and $\omega_B$ are defined by 
$F_{A\phi} = \pa_A A_\phi = -\epsilon_A{}^B B_B$ 
and $d(\hu)_{A\phi}=\pa_A(hu_\phi) = \epsilon_A{}^B \omega_B$, 
respectively, $\epsilon^{AB}$ is an anti-symmetric tensor 
whose signature is defined as $\epsilon^{12} = -1$, and 
$B_\phi$ is defined by $\epsilon_{AB}B_\phi = F_{AB}$.  
Here, $F_{A\phi}$, and $F_{AB}$ are the components of the Faraday 
tensor $\Fabd = (dA)_\albe$, 
$u^\alpha$ is the 4-velocity, $\rho$ the rest mass density and 
$h = (\epsilon+p)/\rho$ the relativistic enthalpy, where $\epsilon$ is the energy density and $p$ is the pressure.  
In the above current, 
$\Lambda$ and $\sqrt{-g}\Lambda_\phi$ are arbitrary functions of 
$\Upsilon$ ($=A_\phi$ for present case) appearing in the first integrals
of  the $t$ and $\phi$ components of the MHD-Euler equation.  
As shown in Eq.~(\ref{eq:current_jphijt_Aphi}), the $t$ and $\phi$ components 
of the 4-current are not independent.  This is a consequence of one of the 
integrability conditions that relate the $A_t$ and $A_\phi$ components to each other, $A_t = A_t(A_\phi)$.  We use the $t$-component 
of Maxwell's equations to determine $j^t$, which is written 
$4\pi\alpha j^t = \dis D_a F^a = \psi^{-6}\zD_a(\psi^6 F^a)$.  
The components $F^a={\gamma^a\!{}_\alpha}F^{\albe}n_\beta$
of the Faraday tensor $\Fabd$ involve the prescribed function 
$A_t(A_\phi)$.  The calculated quantity $j^t$ is then substituted into 
Eq.~(\ref{eq:current_jphijt_Aphi}) to obtain $j^\phi$ for the source, in order 
to integrate the spatial components of  Maxwell's equations.  
It should be noted that, in the perfect MHD case, the current $j^\alpha$ is 
not an independent variable; we introduce $j^\alpha$ as an auxiliary  
variable to derive a formulation suitable for our numerical integration scheme.  

Assuming a one-parameter equation of state $p=p(\rho)$, we choose 
the relativistic enthalpy $h$ as the only independent thermodynamic variable.  
The matter variables $\{h,u^\alpha\}$ are solved respectively from 
a first integral of the MHD-Euler equation, 
the normalization of  4-velocity 
$u_\alpha u^\alpha=-1$, 
the integrability condition of the perfect conductivity condition,   
and the rest mass conservation equation,  as 
\beqn
&&
h\,=\,\Lambda\,(u_t-A'_t u_\phi)^{-1}
\\
&&
u^t = [-\gabd(t^\alpha+v^\alpha)(t^\beta+v^\beta)]^{-1}
\\
&&
u^\phi \,=\, \frac{[\sqrt{-g}\Psi]' B_\phi}{A'_\phi \rho\sqrt{-g}}
\,-\, \frac{A'_t}{A'_\phi}u^t 
%
%(A'_\phi u^\phi + A'_t u^t)\rho\sqrt{-g} 
%\,=\, [\sqrt{-g}\Psi]' B_\phi
\\
&&
u^A = \frac1{\rho\sqrt{-g}}\epsilon^{AB}\pa_B(\sqrt{-g}\Psi).
\label{eq:stream_fn}
\eeqn
%%%and a normalization of the 4 velocity 
%%%$u_\alpha u^\alpha=-1$.  
%%% , where $B_\phi$ is defined from $\epsilon_{AB}B_\phi=F_{AB}=(dA)_{AB}$.  
At each level of the self-consistent iteration, these equations are 
used to update hydrodynamic variables $\{h,u^\alpha\}$ by substituting 
values from the previous iteration level to the right hand sides.  

Finally, to close our formulation for the present calculations, 
we set arbitrary functions following Newtonian studies on 
magnetized rotating stars \cite{YYE06}.  
The functions $A_t(A_\phi)$, $\Lambda(A_\phi)$, 
$[\sqrt{-g}\Lambda_\phi](A_\phi)$, and $[\sqrt{-g}\Psi](A_\phi)$ 
are chosen as follows:
\beqn
&&
A_t(A_\phi) = - \Omega_c A_\phi, 
\label{eq:MHDfnc_At}
\\
&&
\Lambda(A_\phi) = - \Lambda_c A_\phi - {\cal E},
\label{eq:MHDfnc_Lambda}
\\
&&
[\sqrt{-g}\Lambda_\phi](A_\phi) = \frac{a_\Lambda}{k+1}(A_\phi - A_\phi^{\rm max})^{k+1}
\Theta(A_\phi - A_\phi^{\rm max}), \quad 
\label{eq:MHDfnc_Lambda_phi}
\\
&&
[\sqrt{-g}\Psi](A_\phi) = \frac{a_\Psi}{q+1}(A_\phi - A_\phi^{\rm max})^{q+1}
\Theta(A_\phi - A_\phi^{\rm max}), \quad 
\label{eq:MHDfnc_Psi}
\eeqn
where 
$\Omega_c$, $\Lambda_c$, ${\cal E}$ 
$a_\Lambda$, $k$, $a_\Psi$, and $q$ are constants, 
$A^{\rm max}_\phi$ is the maximum value of $A_\phi$ at the stellar surface, 
and $\Theta(x)$ is the Heaviside function.  In the above choice of functions, 
the field $A_\phi$ is always positive.  
The constants $\Lambda_c$, and $a_\Lambda$ are parameters 
given by hand and control the magnetic field strength.  
The constant $a_\Psi$ and the exponent $q$ control the meridional flow, 
where the value of $a_\Psi$ is set to a small value so that the flow 
does not affect the equilibrium of the star, and $q$ is set to one.  
As discussed in \cite{YYE06}, the choice of 
Eq.(\ref{eq:MHDfnc_At}) corresponds to rigid rotation in the limit of 
$a_\Psi \rightarrow 0$ and $B_\phi \rightarrow 0$.  Any other choice 
of the term as a function of $A_\phi$ yields  differential rotation 
in the same limit.  
%%% Also, in the same limit, ${\cal E}$ corresponds to the injection energy.  
In \cite{YYE06}, it is found that 
the solutions have comparable strength in the poloidal and 
toroidal magnetic field components  when 
the index is about $k = 0.1$.  
${\Omega_c}$ and ${\cal E}$ are calculated 
from conditions imposed on the solution, 
which fix the values of central density and axis ratio of 
the deformed star.  

The above set of equations (elliptic equations for the gravitational 
and electromagnetic fields and algebraic relations for the hydrodynamic 
variables) are numerically solved by self-consistent iteration \cite{KEH}.  
In this iteration scheme, the elliptic equations are recast into integral 
equations using the Green function for the flat Laplacian.

The code is developed as  part of the \cocal (Compact Object CALculator) code 
introduced in \cite{cocal}.  
In this code, the Einstein-Maxwell and Maxwell's equations for the fields are solved 
without explicitly imposing axisymmetry.  This allows future extensions of the code  
to compute non-axisymmetric structures such as magnetized NS
with tilted poloidal magnetic fields.  In the numerical solutions 
presented below, the star is covered using 
$161 \times 193 \times 49$  equidistant grid points in 
$(r, \theta, \phi)\in [0,R(\theta,\phi)]\times[0,\pi]\times[0,2\pi]$. The space $\Sigma$ is covered by a computational domain 
$(r, \theta, \phi)\in [0,10^6 R_0]\times[0,\pi]\times[0,2\pi]$ 
with 385 radial non-equidistant grid points and the same equidistant 
$\theta$ and $\phi$ grids, where $R(\theta,\phi)$ represents 
the stellar surface, and $R_0=R(\pi/2,\phi)$ the equatorial radius.  

Further details of the formulation and 
numerical method will be described separately in a forthcoming paper.  

\begin{figure*}
\begin{center}
\includegraphics[height=45mm]{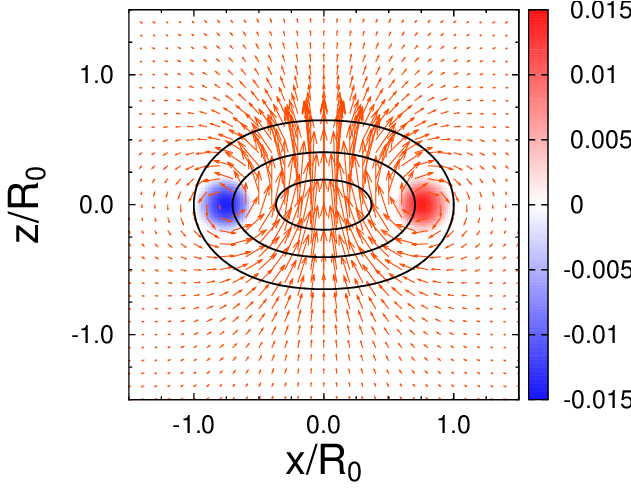}
\includegraphics[height=45mm]{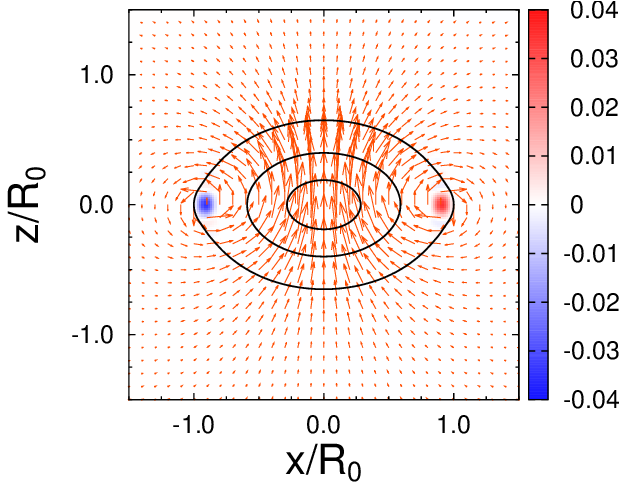}
\includegraphics[height=45mm]{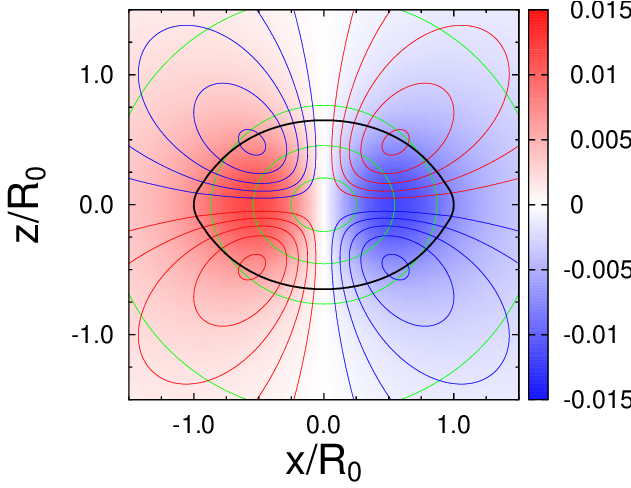}
\caption{Left and Center: Contours of $p/\rho$ (black), density map for $B_\phi$ 
(red and blue), and vector plots for $B_A$ are shown for models I and II, respectively.  
Black contours are drawn at the stellar surface and at $p/\rho=0.05$ and $0.1$.  
Right: Contours for $\psi$ (green), density map for $\tilde \beta_y$ 
(red and blue), and contours for $\tgamma_{xz}$ (blue and red) are shown for  
model II.  
Green contours are drawn in increments of 0.05, starting from 1.2.  
Blue (red) contours correspond to negative (positive) values, drawn in increments of 0.002. The thick black oval corresponds to the  stellar surface.  
The maximum values of the magnetic field components for the solutions of the left and center panels are 
$(B_{\rm pol}^{\rm max}, B_{\rm tor}^{\rm max}) = 
(7.19\times 10^{-2}, 1.32\times 10^{-2})$ and 
$(5.40\times 10^{-2}, 3.21\times 10^{-2})$ 
respectively.  
}
\label{fig:MRNS_contours}
\end{center}
\end{figure*}

\begin{table*}
\begin{tabular}{ccccccccccccc}
\hline
Model & $R_0$ & $R_p/R_0$ & $p/\rho$ & $M_0$ & $\Madm$ & $J/\Madm$ & $T/|W|$ & $\Pi/|W|$ & ${\cal M}/|W|$ & 
${\cal M}_{\rm pol}/|W|$ & ${\cal M}_{\rm tor}/|W|$ & $I_{\rm vir}/|W|$ \\
I&  11.8 &  0.662 &  0.122 &  1.788 &  1.551 &  0.236 &   6.17E-03 &   3.05E-01 &   4.69E-02 &   4.51E-02 &   8.69E-04 &   2.48E-02 \\
II&  12.6 &  0.667 &  0.122 &  1.725 &  1.511 &  0.191 &   4.04E-03 &   3.04E-01 &   5.17E-02 &   4.90E-02 &   2.05E-03 &   2.70E-02 \\
\hline
\end{tabular}
\caption{
Values of physical quantities of selected solutions.  Quantities with dimensions are in 
$G=c=M_\odot=4\pi\varepsilon_0=1$ units (e.g.~$1 M_\odot=145,790$cm).  
$M_0$ is the rest mass and 
$I_{\rm vir}=|2T+3\Pi+{\cal M}+W|$ is the virial integral.  
}
\label{tab:MRNS}
\end{table*}

%%%\section{Numerical solutions}
\paragraph*{\underline{Numerical solutions}:}
We use the relativistic virial theorem for 
perfect fluid Einstein-Maxwell spacetimes derived in \cite{1994CQGra..11..443G}
to study the accuracy of solutions, as well as to quantify 
the amount of electromagnetic energy.  
The relativistic virial theorem is an integral that identically vanishes,  
\beqn
&&
\textstyle
\int_\Sigma \left(T_a{}^a - {\small \frac1{8\pi}} G_a{}^a \right)dV
\nonumber\\
&&\quad
\,=\,
2T \,+\, 3\Pi \,+\, {\cal M} \,+\, W \,+\, \Madm \,-\, \MK \,=\, 0,  \quad
\eeqn
where $G_{ab}$ and $T_{ab}$ are the projections of the Einstein and  stress-energy 
tensors to a hypersurface $\Sigma$.  
For stationary spacetimes, an equality of the ADM mass and the 
Komar mass $\Madm = \MK$ has been established in 
\cite{1978PhLA...69..153B} (also see e.g. \cite{Shibata:2004qz}).  
The quantities $T$, $\Pi$, ${\cal M}$, and $W$ correspond to 
the kinetic, internal, electromagnetic, and gravitational energies 
in the Newtonian limit and are defined by 
\beqn
&& 
\textstyle
T = \frac12\int_\Sigma (\epsilon+p)u_a u^a dV, 
%\\&&
%\textstyle
\qquad
\Pi = \int_\Sigma p\, dV, 
\\&&
\textstyle
{\cal M} = \frac1{16\pi}\int_\Sigma (2 F_a F^a + F_{ab} F^{ab})dV
\\&&
\textstyle
W = \frac1{4\pi}\int_\Sigma [\psi^{-4}(2\tD^a \ln\psi \tD_a \ln\psi - \tD^a \ln\alpha \tD_a \ln\alpha)
\nonumber\\
&&
\textstyle
+\frac34(\Aabd \Aabu -\frac23 K^2)+\frac1{\alpha}K\beta^a\tD_a\ln\alpha+\frac14 \ttR \psi^{-4}]dV, \quad  
\eeqn
where $F_{ab}=\gmaa \gmbb F_\albe$ is the projection of the Faraday tensor onto $\Sigma$.  
$A_{ab}$ and $K$ are the tracefree and trace parts of the extrinsic curvature $K_{ab}$, 
and the tilded quantities $\tD$ and $\ttR$ are associated with the conformal 3-metric $\tgmabd$.  
The magnetic energy term $\cal M$ contains contributions from the poloidal and toroidal 
magnetic fields, for which we respectively define 
\beq 
\textstyle
{\cal M}_{\rm pol}=\frac1{16\pi}\int_\Sigma F_{AB} F^{AB}dV, \quad
{\cal M}_{\rm tor}=\frac1{8\pi}\int_\Sigma F_{A\phi}F^{A\phi}dV. 
\eeq

Quantities including those defined above for 
selected compact star solutions  with mixed 
poloidal and toroidal magnetic fields are listed in Table \ref{tab:MRNS}, 
and corresponding contours in $xz(y=0)$-plane are presented in 
Fig.~\ref{fig:MRNS_contours}.  
In these models, we choose a polytropic equation of state 
$p = K\rho^\Gamma$ with an adiabatic constant $\Gamma=2$, 
and set the parameters in the integrability conditions to 
$(\Lambda_c,a_\Lambda)=(0.8,0.01)$ and $(0.72,0.03)$.  
In spite of the smallness of the ratio $T/|W|$ 
compared to $M/|W|$, the stars are significantly deformed; 
the magnetic fields affect the hydrostatic equilibrium configuration.  
In these models, the maximum values of the poloidal $B$ fields are 
above $10^{18}$ G in cgs Gauss units, 
$(B_{\rm pol}^{\rm max}, B_{\rm tor}^{\rm max}) = 
(1.69\times 10^{18}{\rm G}, 3.11\times 10^{17}{\rm G})$ and 
$(1.27\times 10^{18}{\rm G}, 7.56\times 10^{17}{\rm G})$ 
respectively.

The value of the virial integral $I_{\rm vir}=|2T+3\Pi+{\cal M}+W|$ presented 
in Table \ref{tab:MRNS} indicates 
the deviation of a solution from  magnetic and hydrostationary equilibrium, 
which may be caused either by the numerical error, the time dependence of 
the solution, or both.  Since  stationarity is imposed on the set of 
equations in our formulation, the value should converge to zero as 
the resolution increased.  Due to limited computational resources, 
we haven't performed convergence tests for these magnetized models.\footnote{
We have performed convergence tests of the virial integral 
for relativistic rotating stars turning off the magnetic field 
and reducing the number of grid points.  We have confirmed 2nd order convergnece 
for the solutions calculated with the waveless formulation, as well as 
with the Isenverg-Wilson-Mathews (spatially conformally flat) formulation \cite{Gourg2012,supp2}.
In the result by Kiuchi and Yoshida \cite{Kiuchi:2008ch} for NS with purely 
toroidal magnetic field, in which a finite difference scheme is used, 
a typical value of $I_{\rm vir}/|W|$ is about an order of magnitude smaller than 
our result.  
}
%Instead, we briefly discuss in Appendix \ref{app:virial}  a convergence 
%test result of  the virial integral for relativistic rotating stars with 
%the magnetic field  turned off.  
Nonetheless, the fact 
that the value $I_{\rm vir}/|W|$ is smaller than ${\cal M}/|W|$ suggests 
that  magnetohydrostationary equilibrium is attained in the presented 
solutions.

The magnetic field becomes stronger as the parameter $\Lambda_c$ 
increases.  The magnitude of the toroidal component $B_\phi$ is controlled by 
$a_\Lambda$, but its dependence on $a_\Lambda$ is not monotonic.  
Starting from a small value of $a_\Lambda$, 
$B_\phi$ increases as the value $a_\Lambda$ increases, and $B_\phi$ 
concentrates near the equatorial surface.  
%% 
%% At some point the ratio of toroidal energy to poloidal energy decreases, 
%% so that the toroidal energy can not be as strong as the poloidal energy 
%% in our present model.  
%% 

%%% \section{Discussion}
\paragraph*{\underline{Discussion}:}
The configuration and magnitude of the magnetic fields in the above 
equilibrium solutions largely depend on the choice of arbitrarily 
specifiable functions and associated parameters.  
With our present choice of the functions 
(\ref{eq:MHDfnc_At})-(\ref{eq:MHDfnc_Psi}) and the chosen
range of parameters, the energy of toroidal magnetic 
fields does not exceed that of poloidal fields.  
Such was also the case in previous Newtonian studies 
\cite{YYE06}, to which we have referred in choosing 
the functions (\ref{eq:MHDfnc_At})-(\ref{eq:MHDfnc_Psi}).  
As pointed out in simulations \cite{Braithwaite}, 
the stable magnetized configurations may consist of nearly equal 
magnitude of poloidal and toroidal magnetic fields.  
Some  recent studies have succeeded 
in constructing such models in the slow rotation and weak magnetic field limit 
\cite{Yoshida:2012mu,Ciolfi:2013dta}
or in the Newtonian limit \cite{Fujisawa:2013kxa}.  
It will  certainly be possible to find solutions with the toroidal energy 
nearly equal to, or stronger than, the poloidal energy in our relativistic 
calculations by surveying a variety of the arbitrary functions and 
their large parameter space.  
We will perform such systematic calculations elsewhere.

In our formulation for the Einstein-Maxwell and Maxwell's 
equations, we only set the time derivatives to zero and recast them 
in elliptic equations.  This does not guarantee 
a solution to be stationary in general.  
Imposition of symmetries with respect 
to vectors $t^\alpha$ and $\phi^\alpha$ on the set of ideal 
MHD equations guarantees that the solutions are exact 
stationary and axisymmetric ones.  
For more interesting nonaxisymmetric quasiequilibrium data, 
such as a magnetized NS whose magnetic axis is tilted with 
respect to the rotation axis or a magnetized BNS, a part of 
the present code for computing the metric and electromagnetic 
field can be used as it is.\footnote{
It is necessary to develop another 
formulation, such as \cite{Bekenstein:2006xh,Uryu:2010su}, 
and routines to solve 
it for computing magnetized quasiequilibriums.}  
As in our previous work for BNS \cite{WLBNS}, such a computation 
reproduces initial data in which the radiation content is discarded.  
With a rather minor extension of the code, we could also compute 
solutions associated with stationary radiations.

The solutions presented in this paper are useful as initial data 
sets for numerical simulations \cite{GRsimulations}, as well as 
unperturbed states for linear stability analysis 
(see e.g.~\cite{Yoshida:2012rk}).  
The search for stable and realistic 
magnetized neutron star models is the next step of 
this work.

\section*{Acknowledgements}
%\paragraph*{\underline{Acknowledgement}:}
%
%
We gratefully acknowledge support under 
JSPS Grant-in-Aid for Scientific Research(C) 23540314 and 25400262, 
Grant-in-Aid for Scientific Research on Innovative Areas 24103006, 
DFG grant SFB/Transregio 7 ``Gravitational Wave Astronomy''
and STFC grant PP/E001025/1.
%%%, and MEXT Grant-in-Aid for Scientific Research
%%% on Innovative Area 20105004.  
%


\begin{thebibliography}{99}

%\cite{Duncan:1992hi}
\bibitem{Duncan:1992hi} 
  R.~C.~Duncan and C.~Thompson,
  %``Formation of very strongly magnetized neutron stars - implications for gamma-ray bursts,''
  Astrophys.\ J.\  {\bf 392}, L9 (1992).
  %%CITATION = ASJOA,392,L9;%%
  %823 citations counted in INSPIRE as of 07 May 2013

%\cite{Metzger:2010pp}
\bibitem{Metzger:2010pp} 
  B.~D.~Metzger, D.~Giannios, T.~A.~Thompson, N.~Bucciantini and E.~Quataert,
   Mon.\ Not.\ Roy.\ Astron.\ Soc.\  {\bf 413}, 2031 (2011)
  %``The Proto-Magnetar Model for Gamma-Ray Bursts,''
  %arXiv:1012.0001 [astro-ph.HE].
  %%CITATION = ARXIV:1012.0001;%%
  %19 citations counted in INSPIRE as of 08 May 2013


%\cite{Bocquet:1995je}
\bibitem{Bocquet:1995je} 
  M.~Bocquet, S.~Bonazzola, E.~Gourgoulhon and J.~Novak,
  %``Rotating neutron star models with magnetic field,''
  Astron.\ Astrophys.\  {\bf 301}, 757 (1995)
  %[gr-qc/9503044].
  %%CITATION = GR-QC/9503044;%%
  %128 citations counted in INSPIRE as of 07 May 2013

%\cite{Kiuchi:2008ch}
\bibitem{Kiuchi:2008ch} 
  K.~Kiuchi and S.~Yoshida,
  %``Relativistic stars with purely toroidal magnetic fields,''
  Phys.\ Rev.\ D {\bf 78}, 044045 (2008)
%  [arXiv:0802.2983 [astro-ph]].
  %%CITATION = ARXIV:0802.2983;%%
  %16 citations counted in INSPIRE as of 07 May 2013

%\cite{Frieben:2012dz}
\bibitem{Frieben:2012dz} 
  J.~Frieben and L.~Rezzolla,
  %``Equilibrium models of relativistic stars with a toroidal magnetic field,''
  Mon.\ Not.\ Roy.\ Astron.\ Soc.\  {\bf 427}, 3406 (2012)
%  [arXiv:1207.4035 [gr-qc]].
  %%CITATION = ARXIV:1207.4035;%%
  %4 citations counted in INSPIRE as of 07 May 2013

%\bibitem{circular}
%see e.g., 
\bibitem{Carter73}
B.~Carter, in \emph{Black holes --- Les Houches 1972}, edited by C.~DeWitt
\& B.S.~DeWitt, Gordon and Breach, New York (1973), p.~57; 

%\cite{Oron:2002gs}
\bibitem{Oron:2002gs} 
  A.~Oron,
  %``Relativistic magnetized star with poloidal and toroidal fields,''
  Phys.\ Rev.\ D {\bf 66}, 023006 (2002);
  %%CITATION = PHRVA,D66,023006;%%
  %13 citations counted in INSPIRE as of 27 Sep 2013, 

\bibitem[Friedman \& Stergioulas(2013)]{2013rrs..book.....F} 
J.~L.~Friedman and N.~Stergioulas, 
\emph{Rotating Relativistic Stars}, 
Cambridge University Press, Cambridge, UK, 2013.

\bibitem[Straumann(2013)]{2013gere.book.....S} 
N.~Straumann, \emph{General Relativity}, Springer 
Science+Business Media Dordrecht, 2013.  


\bibitem[Bekenstein \& Oron(1978)]{1978PhRvD..18.1809B} 
J.~D.~Bekenstein and E.~Oron,
\prd, 18, 1809 (1978);
%

\bibitem{IokaSasaki} 
%\cite{Ioka:2003dd}
%\bibitem{Ioka:2003dd} 
  K.~Ioka and M.~Sasaki,
  %``Grad-Shafranov equation in noncircular stationary axisymmetric space-times,''
  Phys.\ Rev.\ D {\bf 67}, 124026 (2003);
%  [gr-qc/0302106].
  %%CITATION = GR-QC/0302106;%%
  %15 citations counted in INSPIRE as of 07 May 2013
%\cite{Ioka:2003nh}
%\bibitem{Ioka:2003nh} 
  K.~Ioka and M.~Sasaki,
  %``Relativistic stars with poloidal and toroidal magnetic fields and meridional flow,''
  Astrophys.\ J.\  {\bf 600}, 296 (2004)
%  [astro-ph/0305352].
  %%CITATION = ASTRO-PH/0305352;%%
  %28 citations counted in INSPIRE as of 07 May 2013

\bibitem{Ciolfi}
%\cite{Ciolfi:2009bv}
%\bibitem{Ciolfi:2009bv} 
  R.~Ciolfi, V.~Ferrari, L.~Gualtieri and J.~A.~Pons,
  %``Relativistic models of magnetars: the twisted-torus magnetic field configuration,''
  Mon.\ Not.\ Roy.\ Astron.\ Soc.\  {\bf 397}, 913 (2009); 
%  [arXiv:0903.0556 [astro-ph.SR]].
  %%CITATION = ARXIV:0903.0556;%%
  %26 citations counted in INSPIRE as of 01 Jul 2013
%\cite{Ciolfi:2010td}
%\bibitem{Ciolfi:2010td} 
  R.~Ciolfi, V.~Ferrari and L.~Gualtieri,
  %``Structure and deformations of strongly magnetized neutron stars with twisted torus configurations,''
  Mon.\ Not.\ Roy.\ Astron.\ Soc.\  {\bf 406}, 2540 (2010)
%  [arXiv:1003.2148 [astro-ph.SR]].
  %%CITATION = ARXIV:1003.2148;%%
  %20 citations counted in INSPIRE as of 01 Jul 2013


%\cite{Yoshida:2012mu}
\bibitem{Yoshida:2012mu} 
  S.~Yoshida, K.~Kiuchi and M.~Shibata,
  %``Stably stratified magnetized stars in general relativity,''
  Phys.\ Rev.\ D {\bf 86}, 044012 (2012)
%  [arXiv:1207.1942 [gr-qc]].
  %%CITATION = ARXIV:1207.1942;%%
  %2 citations counted in INSPIRE as of 07 May 2013

%\cite{Pili:2014npa}
 \bibitem{Pili:2014npa}
        A.~G.~Pili, N.~Bucciantini and L.~Del Zanna,
        Mon.\ Not.\ Roy.\ Astron.\ Soc.,\ {\bf 439}, 3541 (2014)

%\cite{Gourgoulhon:2011gz}
\bibitem{Gourgoulhon:2011gz} 
  E.~Gourgoulhon, C.~Markakis, K.~Uryu and Y.~Eriguchi,
  %``Magnetohydrodynamics in stationary and axisymmetric spacetimes: a fully covariant approach,''
  Phys.\ Rev.\ D {\bf 83}, 104007 (2011)
%  [arXiv:1101.3497 [gr-qc]].
  %%CITATION = ARXIV:1101.3497;%%
  %4 citations counted in INSPIRE as of 07 May 2013

\bibitem{Braithwaite} 
%\cite{Braithwaite:2005ps}
%\bibitem{Braithwaite:2005ps} 
  J.~Braithwaite and H.~C.~Spruit,
  %``Structure of the magnetic fields in A stars and white dwarfs,''
  Nature {\bf 431}, 819 (2004);
%  [astro-ph/0502043].
  %%CITATION = ASTRO-PH/0502043;%%
  %105 citations counted in INSPIRE as of 08 May 2013
%\bibitem[Braithwaite \& Nordlund(2006)]{2006A&A...450.1077B}
J.~Braithwaite and {\AA}.~Nordlund, A\&A 450, 1077 (2006);
J.~Braithwaite, A\&A \textbf{453}, 687 (2006);
J.~Braithwaite, A\&A \textbf{469}, 275 (2007) 
J.~Braithwaite, MNRAS \textbf{397} 763 (2009);
V.~Duez, J.~Braithwaite, and S. Mathis, ApJ {\bf{724}} L34 (2010).

\bibitem{GRsimulations}
%\cite{Shibata:2005gp}
%\bibitem{Shibata:2005gp} 
  M.~Shibata and Y.~-i.~Sekiguchi,
  %``Magnetohydrodynamics in full general relativity: Formulation and tests,''
  Phys.\ Rev.\ D {\bf 72}, 044014 (2005);
%  [astro-ph/0507383].
  %%CITATION = ASTRO-PH/0507383;%%
  %57 citations counted in INSPIRE as of 10 May 2013
%
%\cite{Duez:2006qe}
%\bibitem{Duez:2006qe}
  M.~D.~Duez, Y.~T.~Liu, S.~L.~Shapiro, M.~Shibata and B.~C.~Stephens, 
  %``Evolution of magnetized, differentially rotating neutron stars: Simulations in full general relativity,''
  Phys.\ Rev.\ D {\bf 73} 104015, (2006);
%  [astro-ph/0605331].
  %%CITATION = ASTRO-PH/0605331;%%
  %58 citations counted in INSPIRE as of 10 May 2013
%
%\cite{Giacomazzo:2007ti}
%\bibitem{Giacomazzo:2007ti} 
  B.~Giacomazzo and L.~Rezzolla,
  %``WhiskyMHD: A New numerical code for general relativistic magnetohydrodynamics,''
  Class.\ Quant.\ Grav.\  {\bf 24}, S235 (2007);
%  [gr-qc/0701109].
  %%CITATION = GR-QC/0701109;%%
  %27 citations counted in INSPIRE as of 10 May 2013
%
%\bibitem[Shibata et al.(2011)]{2011ApJ...734L..36S} 
M.~Shibata, Y.~Suwa, K.~Kiuchi, and K.~Ioka, ApJL 734, L36 (2011);
%
%\cite{Liebling:2010bn}
%\bibitem{Liebling:2010bn} 
  S.~L.~Liebling, L.~Lehner, D.~Neilsen and C.~Palenzuela,
  %``Evolutions of Magnetized and Rotating Neutron Stars,''
  Phys.\ Rev.\ D {\bf 81}, 124023 (2010);
%  [arXiv:1001.0575 [gr-qc]].
  %%CITATION = ARXIV:1001.0575;%%
  %13 citations counted in INSPIRE as of 10 May 2013
%
%\cite{Kuroda:2010yc}
%\bibitem{Kuroda:2010yc} 
  T.~Kuroda and H.~Umeda,
  %``Three Dimensional Magneto Hydrodynamical Simulations of Gravitational Collapse of a 15Msun Star,''
  Astrophys.\ J.\ Suppl.\  {\bf 191}, 439 (2010);
%  [arXiv:1008.1370 [astro-ph.SR]].
  %%CITATION = ARXIV:1008.1370;%%
  %7 citations counted in INSPIRE as of 10 May 2013
%
%\cite{Kiuchi:2011yt}
%\bibitem{Kiuchi:2011yt} 
  K.~Kiuchi, S.~Yoshida and M.~Shibata,
  %``Non-axisymmetric instabilities of neutron star with toroidal magnetic fields,''
  Astron.\ Astrophys.\  {\bf 532}, A30 (2011);
%  [arXiv:1104.5561 [astro-ph.HE]].
  %%CITATION = ARXIV:1104.5561;%%
  %4 citations counted in INSPIRE as of 10 May 2013
%
%\cite{Kiuchi:2012qv}
%\bibitem{Kiuchi:2012qv} 
  K.~Kiuchi, K.~Kyutoku and M.~Shibata,
  %``Three dimensional evolution of differentially rotating magnetized neutron stars,''
  Phys.\ Rev.\ D {\bf 86}, 064008 (2012);
%  [arXiv:1207.6444 [astro-ph.HE]].
  %%CITATION = ARXIV:1207.6444;%%
  %2 citations counted in INSPIRE as of 10 May 2013
%
%\cite{Etienne:2011re}
%\bibitem{Etienne:2011re} 
  Z.~B.~Etienne, V.~Paschalidis, Y.~T.~Liu and S.~L.~Shapiro,
  %``Relativistic MHD in dynamical spacetimes: Improved EM gauge condition for AMR grids,''
  Phys.\ Rev.\ D {\bf 85}, 024013 (2012)
%  [arXiv:1110.4633 [astro-ph.HE]].
  %%CITATION = ARXIV:1110.4633;%%
  %11 citations counted in INSPIRE as of 10 May 2013  

%\cite{Yoshida:2012rk}
\bibitem{Yoshida:2012rk} 
  S.~Yoshida,
  %``Non-axisymmetric oscillations of rapidly rotating relativistic stars by conformal flatness approximation,''
  Phys.\ Rev.\ D {\bf 86}, 104055 (2012)
%  [arXiv:1205.3893 [gr-qc]].
  %%CITATION = ARXIV:1205.3893;%%


%\cite{}
\bibitem{Papapetrou} 
  A.~Papapetrou,
  %``Champs gravitationnels stationnaires a symetrie axiale,''
  Ann.\ Inst.\ Henri Poincare Phys.\ Theor.\ IV {\bf }, 83 (1966).
  %%CITATION = AIPTE,IV,83;%%
  %2 citations counted in INSPIRE as of 24 Nov 2013

%\cite{}
\bibitem{Kundt} 
  W.~Kundt and M.~Trumper,
  %``Orthogonal decomposition of axi-symmetric stationary spacetimes,''
  Z.\ Phys.\  {\bf 192}, 419 (1966).
  %%CITATION = ZEPYA,192,419;%%
  %33 citations counted in INSPIRE as of 24 Nov 2013

\bibitem[Chandrasekhar(1992)]{1992mtbh.book.....C} S.~Chandrasekhar, 
\emph{The Mathematical Theory of Black Holes},  
New York : Oxford University Press, 1992.  

\bibitem[Wald(1984)]{1984ucp..book.....W} 
R.~M.~Wald, \emph{General Relativity}, 
University of Chicago Press, Chicago, 1984.  

\bibitem{supp1}
See Supplemental Material at [URL] for a 
brief review on stationary and axisymmetric
spacetimes for rotating and magnetized compact
stars.

%\cite{Birkl:2010hc}
\bibitem{Birkl:2010hc} 
  R.~Birkl, N.~Stergioulas and E.~Muller,
  %``Stationary, Axisymmetric Neutron Stars with Meridional Circulation in General Relativity,''  
  Phys.\ Rev.\ D {\bf 84}, 023003 (2011)  
  %[arXiv:1011.5475 [gr-qc]].  
  %%CITATION = ARXIV:1011.5475;%%  
  %1 citations counted in INSPIRE as of 23 May 2013

\bibitem[Gourgoulhon 
\& Bonazzola(1993)]{1993PhRvD..48.2635G} 
E.~Gourgoulhon, and S.~Bonazzola, \prd, 48, 2635 (1993) 

%\cite{Shibata:2004qz}
\bibitem{Shibata:2004qz} 
  M.~Shibata, K.~Uryu and J.~L.~Friedman,
  %``Deriving formulations for numerical computation of binary neutron stars in quasicircular orbits,''
  Phys.\ Rev.\ D {\bf 70}, 044044 (2004)
  [Erratum-ibid.\ D {\bf 70}, 129901 (2004)]
%  [gr-qc/0407036].
  %%CITATION = GR-QC/0407036;%%
  %27 citations counted in INSPIRE as of 07 May 2013

\bibitem{WLBNS}
%\cite{Uryu:2005vv}
%\bibitem{Uryu:2005vv} 
  K.~Uryu, F.~Limousin, J.~L.~Friedman, E.~Gourgoulhon and M.~Shibata,
  %``Binary neutron stars in a waveless approximation,''
  Phys.\ Rev.\ Lett.\  {\bf 97}, 171101 (2006);
%  [gr-qc/0511136].
  %%CITATION = GR-QC/0511136;%%
  %32 citations counted in INSPIRE as of 07 May 2013
%\cite{Uryu:2009ye}
%\bibitem{Uryu:2009ye} 
  K.~Uryu, F.~Limousin, J.~L.~Friedman, E.~Gourgoulhon and M.~Shibata,
  %``Non-conformally flat initial data for binary compact objects,''
  Phys.\ Rev.\ D {\bf 80}, 124004 (2009).
%  [arXiv:0908.0579 [gr-qc]].
  %%CITATION = ARXIV:0908.0579;%%
  %18 citations counted in INSPIRE as of 07 May 2013

\bibitem{Gourg2012}
E.~Gourgoulhon : \emph{3+1 Formalism in General Relativity; Bases of Numerical Relativity},
Springer (Berlin) (2012) 

%\cite{Bonazzola:2003dm}
\bibitem{Bonazzola:2003dm} 
  S.~Bonazzola, E.~Gourgoulhon, P.~Grandclement and J.~Novak,
  %``A Constrained scheme for Einstein equations based on Dirac gauge and spherical coordinates,''
  Phys.\ Rev.\ D {\bf 70}, 104007 (2004)
  [gr-qc/0307082].
  %%CITATION = GR-QC/0307082;%%
  %65 citations counted in INSPIRE as of 07 May 2013


\bibitem{YYE06}
%\bibitem[Yoshida \& Eriguchi(2006)]{2006ApJS..164..156Y} 
S.~Yoshida, and Y.~Eriguchi, ApJS, 164, 156 (2006); 
%\bibitem[Yoshida et al.(2006)]{2006ApJ...651..462Y} 
S.~Yoshida, S.~Yoshida, and Y.~Eriguchi, \apj, 651, 462 (2006)


\bibitem{KEH}
%\bibitem{OM68}
J.~P.~Ostriker, and J.~W.-K.~Mark, \apj 151, 1075 (1968);
%%
%%\bibitem{Hachisu86}
I.~Hachisu, ApJS 62, 461 (1986); ibid. 61, 479 (1986);
%%
%%\bibitem{KEH89}
H.~Komatsu, Y.~Eriguchi, and I.~Hachisu, 
Mon.\ Not.\ Roy.\ Astron.\ Soc.\ 237, 355 (1989) 


\bibitem{cocal} 
%\cite{Huang:2008vp}
%\bibitem{Huang:2008vp} 
  X.~Huang, C.~Markakis, N.~Sugiyama and K.~Uryu,
  %``Quasi-equilibrium models for triaxially deformed rotating compact stars,''  
  Phys.\ Rev.\ D {\bf 78}, 124023 (2008)  
  %[arXiv:0809.0673 [astro-ph]].  
  %%CITATION = ARXIV:0809.0673;%%  
  %3 citations counted in INSPIRE as of 23 May 2013
%
%\cite{Uryu:2011ky}
%\bibitem{Uryu:2011ky} 
  K.~Uryu and A.~Tsokaros,
  %``A new code for equilibriums and quasiequilibrium initial data of compact objects,''  
  Phys.\ Rev.\ D {\bf 85}, 064014 (2012)  
  %%%[arXiv:1108.3065 [gr-qc]].  
  %%CITATION = ARXIV:1108.3065;%%  %4 citations counted in INSPIRE as of 23 May 2013
%\cite{Uryu:2012uh}
%\bibitem{Uryu:2012uh} 
  K.~Uryu, A.~Tsokaros and P.~Grandclement,
  %``New code for equilibriums and quasiequilibrium initial data of compact objects. II. Convergence tests and comparisons of binary black hole initial data,''  
  Phys.\ Rev.\ D {\bf 86}, 104001 (2012)  
  %%%[arXiv:1210.5811 [gr-qc]].  
  %%CITATION = ARXIV:1210.5811;%%


\bibitem[Gourgoulhon \& Bonazzola(1994)]{1994CQGra..11..443G} 
E.~Gourgoulhon, and S.~Bonazzola, Classical and Quantum Gravity, 11, 443 (1994)

\bibitem[Beig(1978)]{1978PhLA...69..153B} 
R.~Beig, Physics Letters A, 69, 153 (1978)


\bibitem{supp2} 
See Supplemental Material at [URL] for 
a convergence test of \cocal code on 
the virial integral for relativistic 
rotating stars.  


%\cite{Bekenstein:2006xh}
\bibitem{Bekenstein:2006xh} 
  J.~D.~Bekenstein and G.~Betschart,
  %``Perfect magnetohydrodynamics as a field theory,''
  Phys.\ Rev.\ D {\bf 74}, 083009 (2006)
%  [gr-qc/0608053].
  %%CITATION = GR-QC/0608053;%%


%\cite{Uryu:2010su}
\bibitem{Uryu:2010su} 
  K.~Uryu, E.~Gourgoulhon and C.~Markakis,
  %``Thermodynamics of magnetized binary compact objects,''
  Phys.\ Rev.\ D {\bf 82}, 104054 (2010)
%  [arXiv:1010.4409 [gr-qc]].
  %%CITATION = ARXIV:1010.4409;%%
  %5 citations counted in INSPIRE as of 10 May 2013



%\cite{Ciolfi:2013dta}
\bibitem{Ciolfi:2013dta} 
  R.~Ciolfi and L.~Rezzolla,
  %``Twisted-torus configurations with large toroidal magnetic fields in relativistic stars,''
  Mon.\ Not.\ Roy.\ Astron.\ Soc.,\ {\bf 435}, L43 (2013)
  %arXiv:1306.2803 [astro-ph.SR].
  %%CITATION = ARXIV:1306.2803;%%

%\cite{Fujisawa:2013kxa}
\bibitem{Fujisawa:2013kxa} 
  K.~Fujisawa and Y.~Eriguchi,
  %``Coexistence of oppositely flowing multi-$\varphi$-currents: Key to large toroidal magnetic fields within stars,''
  Mon.\ Not.\ Roy.\ Astron.\ Soc.,\ {\bf 432}, 1245 (2013)
%  [arXiv:1304.0549 [astro-ph.HE].
  %%CITATION = ARXIV:1304.0549;%%


\end{thebibliography}
\end{document}